\documentclass[a4paper,12pt,oneside,dvips,driver]{article}
\usepackage[english]{babel}
\usepackage{pstricks,pst-node,pst-coil}
\usepackage{epsfig}
\usepackage{amscd}
\usepackage{amssymb}
\usepackage{latexsym}
\usepackage{psfrag}

\newcommand{\beq}{\begin{equation}}
\newcommand{\eeq}{\end{equation}}
\newcommand{\bdm}{\begin{displaymath}}
\newcommand{\edm}{\end{displaymath}}
\newcommand{\be}{\beta}
\newcommand{\al}{\alpha}

\begin{document}

\author{P. Jakubczyk  and  M. Napi\'{o}rkowski\\ Instytut Fizyki
Teoretycznej, Uniwersytet Warszawski  \\  00-681 Warszawa, Ho\.za  69,
Poland } 
\title{Interfacial correlation  function for adsorption on a
disc} 
\date{} 
\maketitle 
\begin{abstract}

{Within the Landau-Ginzburg-Wilson approach we derive the effective  Hamiltonian 
governing fluctuations of an interface between two phases, one of them adsorbed on a 
disclike substrate. For large disc radii and for temperatures close to the 
wetting temperature of the corresponding planar substrate the 
expressions for the height-height correlation function and the accompanying 
correlation angle are derived. Their dependence on the disc radius and the reduced 
temperature is discussed and the relevant scaling regimes are identified.} \\


\noindent{\noindent Corresponding author: Pawe\l \quad Jakubczyk Instytut Fizyki Teoretycznej, Uniwersytet Warszawski 0-681 Warszawa, Ho\.za 69, Poland; tel. 0048225532317, e-mail: Pawel.Jakubczyk@fuw.edu.pl}

\end{abstract}
\newpage
\section{Introduction}
\renewcommand{\theequation}{1.\arabic{equation}} 
\setcounter{equation}{0}
\vspace*{0.5cm}

The presence of a substrate favouring one of two coexisting bulk phases may cause 
this phase to form 
 a film adsorbed at its surface. At a certain temperature $T_W$ below the bulk critical 
temperature $T_C$ 
the system may undergo a wetting transition at which the thickness of the film 
becomes macroscopic. Such wetting transitions have been extensively studied during the 
last years, especially for the planar substrates; for reviews see [1-3]. In addition, a 
lot of effort has 
 been made towards understanding the modifications of system's wetting behaviour introduced 
by structured substrates, including chemically homogeneous nonplanar substrates [4-14]. 
In particular, for spherical and  cylindrical geometries [15-25] the substrate's finite 
and positive radius $R$ prevents the divergence 
of the adsorbed film's thickness so that wetting transition point may be reached 
only in the limiting case
 $R = \infty$. 
This is due to the fact that - unlike in the planar case - the area of the interface is not 
invariant under 
the change of its position - and increases with the increasing layer's thickness.  The 
accompanying growth of the interfacial free energy prevents the wetting transition. 

To our knowledge not much attention has been  paid to studying interfacial 
correlation functions in such systems, especially from the point of view of their 
temperature and substrate's radius depencence \cite{Gil}. Our first aim is to 
systematically derive the effective interfacial Hamiltonian 
for adsorption on a disc including its radius and temperature dependence. 
This Hamiltonian is then  used to evaluate the interfacial correlation function 
and the corresponding 
correlation angle parametrized by $R$ and $T$. In particular, one is interested 
in seeing how the divergence 
of the correlation angle close to $T_{W}$ is prevented by finite values of $R$. \\

In Section 2 we introduce the Landau-Ginzburg-Wilson theory for a 2-dimensional system 
consisting of two phases 
coexisting in presence of a disclike substrate. This theory serves to derive within 
the mean field approach the effective 
interface Hamiltonian. We discuss the curvature-induced modifications of its form 
as compared to the planar case and propose a perturbative 
procedure yielding approximate expression for 
the equilibrium interface position at a given thermodynamic state. 

Section 3 contains the discussion of the correlation function for an interface 
fluctuating around its 
mean circular shape. Within the Gaussian approximation the expressions 
for the correlation angle and the amplitude of the correlation function are derived. 

In Section 4 we apply the specific results obtained in Section 2 to 
evaluate the correlation function. After 
introducing natural scaling variables we discuss the asymptotic behaviour of 
the correlation length and identify 
the different scaling regimes. In Section 5 we summarize our results. 

\section{The Hamiltonian}
\renewcommand{\theequation}{2.\arabic{equation}} 
\setcounter{equation}{0}
\vspace*{0.5cm}

The standard method of deriving the effective Hamiltonians amounts to 
defining a new set of variables describing the system's state on the 
coarse-grained scale and 
introducing constraints which specify a restiction on the allowed set 
of microstates compatible 
with fixed values of new variables. The effective Hamiltonian is then 
extracted from the logarithm of statistical 
sum performed under this constraint. The mean field version of this procedure 
amounts to 
minimization of the microscopic Hamiltonian under the appropriate constraint. 

Our derivation of the effective Hamiltonian for a 1-dimensional interface fluctuating 
in presence of a disclike substrate follows the lines first outlined in 
Ref.\cite{Jin} for the planar case. It starts from the description 
of a two-dimensional uniaxial ferromagnet  based on the 
Landau-Ginzburg-Wilson Hamiltonian 

\beq
\label{LGW}
\mathcal{H}[m]=\int_{0}^{2\pi} d\phi\Bigg[\int_{R}^{\infty}rdr\Big[\frac{K}{2}(\nabla m)^2+
\Phi(m)\Big]+R\,\Phi_1(m\mid_{r=R})\Bigg] \quad, 
\eeq
where $m(\vec{r})=m(r,\phi)$ is a scalar order parameter (magnetization), $R$ - the 
substrate's radius, 
 $\Phi(m)$ and $\Phi_1(m|_{r=R})$ - the bulk and surface free energy densities. 
The bulk free energy density 
at coexistence $\Phi_0(m)$ exhibits two equally deep minima located at $m_{\alpha 0}<0$ 
and $m_{\beta 0}>0$ corresponding to two coexisting phases with opposite magnetizations. 
In our approach  $\Phi_0(m)$ has the form of double parabola \cite{LipowskyDP}
\beq
\label{Phi_0}
\Phi_0(m)=\left\{ \begin{array}{ll}
\frac{Kp_{\alpha}^2}{2}(m-m_{\alpha 0})^2\,, & \textrm{for $m\leq m^*$}\\
\frac{Kp_{\beta}^2}{2}(m-m_{\beta 0})^2\,, & \textrm{for $m \geq m^*$},
\end{array} \right.
\eeq
where $K>0$, and $p_{\gamma}^{-1}=\xi_{\gamma}$ is the bulk correlation length 
in phase $\gamma$, $\gamma=\alpha, \beta$. The parameter $m^{*}$ determines 
the value of the order parameter $m$ at which the $\beta$-phase goes over into the 
$\alpha$-phase, i.e., it fixes the position of the $\alpha-\beta$ interface. The choice 
of $m^*$ value is arbitrary \cite{Backx}, \cite{ParryM}; in this paper 
it is set equal to zero \cite{Jin}. 
The continuity of $\Phi_0$ at $m=0$ implies the relation 
$\frac{p_{\alpha}m_{\alpha 0}}{p_{\beta}m_{\beta 0}}=-1$ between the model parameters. 
The double parabola approximation 
to the general $m^4$ model makes further calculations tractable as 
it yields a linear equation for the constrained magnetization profile. 
The cost to be paid for this simplification is the necessity to limit the considered system's states to these far from bulk criticality ($T \ll T_C$) (see \cite{LipowskyDP}, \cite{Jin}). For macroscopic disc radii this corresponds to $p_\gamma R\gg 1$ and $\xi_\gamma$ not too large as compared to the surface enhancement length $\frac{K}{|g|}$ \cite{Upton}. The opposite case was considered in Refs[18-19]. 

In the presence of external magnetic field $h$ the free energy density 
$\Phi(m)=\Phi_{0}(m) - h\,m$ may be rewritten as 

\beq
\label{Phi}
\Phi (m)=\left\{ \begin{array}{ll}
\frac{Kp_{\alpha}^2}{2}(m-m_{\alpha 0 h})^2+\Phi_{\alpha}& \textrm{for $m\leq 0$}\\ 
\frac{Kp_{\beta}^2}{2}(m-m_{\beta 0 h})^2+\Phi_{\beta} & \textrm{for $m \geq 0$}, 
\end{array} \right. 
\eeq
where $m_{\gamma 0 h}=m_{\gamma 0}+\frac{h}{Kp_{\gamma}^2}$ and 
$\Phi_{\gamma} = - h\,(m_{\gamma 0}+\frac{h}{2Kp_{\gamma}^2})$ 
are the  equilibrium bulk magnetization and free energy density 
of phase $\gamma$, respectively. Note that 
for $h=0$ one has $\Phi_{\gamma}=0$. From now on we consider the thermodynamic states 
of the system corresponding to bulk coexistence; more precisely we set $h=0^{-}$ to 
stabilize the $\alpha$ phase in the bulk.  \\
The surface free energy density is reperesented by the expression 

\beq 
\label{Phi1}
\Phi_1(m_1)=-h_1m_1-\frac{1}{2}gm_1^2,
\eeq
where $m_1=m(r=R,\phi)$, $h_1>0$ is the surface magnetic field favouring the phase 
$\beta$ at the substrate, and $g$ is the surface enhancement parameter. In the present 
analysis we assume 
$g<-Kp_{\beta}$ which corresponds to a second order wetting transition 
of a reference planar system \cite{Jin}. \\
 
\begin{figure}
\begin{center}
\includegraphics[width=0.5\textwidth]{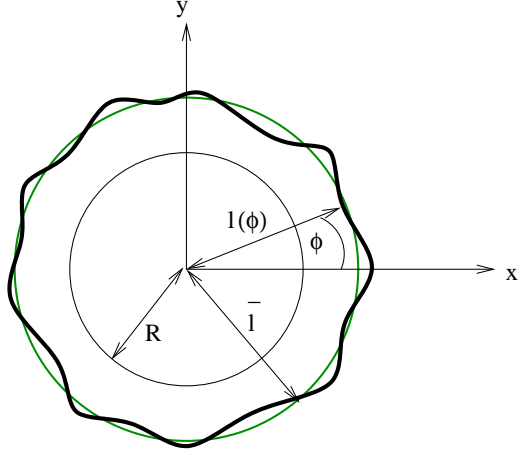}
\caption{A schematic illustration of an interface fluctuating around its 
equilibrium position 
$\bar{l}$ in presence of a disclike substrate with radius $R$.}  
\end{center}
\end{figure}
Minimization of Eq.(\ref{LGW}) is performed under the crossing constraint  
$m(r=l(\phi),\phi)=0$,  
where $r = l(\phi) > R$ specifies the position of the interface. This procedure 
yields the Helmholtz equation for the constrained equilibrium magnetization 
$\widetilde{m}$ 
\beq
K\Delta \widetilde{m}=\frac{d\Phi}{d\widetilde{m}} 
\eeq
which has to be solved with the boundary conditions 
\beq
\label{boundary}
\left.K\frac{\partial \widetilde{m}}{\partial r}\right|_{r=R} \,= \,
\frac{d\Phi_1( \widetilde{m} _1)}{d \widetilde{m} _1} \quad,
\eeq
$\lim_{r\to\infty} \widetilde{m}(r,l_{0}) = m_{\alpha 0 }$, and 
$\lim_{r\to l_{0}^-}\widetilde{m}(r,l_{0}) = 
\lim_{r\to l_{0}^+}\widetilde{m}(r,l_{0})=0$. \\

In the radially symmetric case, i.e., when $l=l_{0}=const$ the constrained 
 magnetization profile $\widetilde{m}_{0}(r,l_{0})$ 
fulfils the zeroth order modified Bessel equation.  Its general solution has 
the following form 
\beq 
\widetilde{m}_{0}(r,l_{0}) - m_{\gamma 0 } = 
C_{\gamma}K_0(p_{\gamma}r)+D_{\gamma}I_0(p_{\gamma}r), \label{m(r)} 
\eeq
where $K_0$ and $I_0$ are the $0$-th order modified Bessel functions, while 
$\gamma =\alpha$ corresponds to $r\geq l$, and $\gamma=\beta$ to 
$r\leq l$, respectively. 
The constants $C_{\gamma}$, $D_{\gamma}$ are determined 
from the above boundary conditions: 
\beq
\label{Const1}
C_{\alpha}=\frac{-m_{\alpha 0 }}{K_0(p_{\alpha}l_{0})}, \quad 
D_{\alpha}=0 \quad,
\eeq

\beq
\label{Const2}
C_{\beta}=\Big[\frac{-m_{\beta 0}}{I_0(p_\beta l_{0})} + 
F_1(R)\Big]\Big[\frac{K_0(p_{\beta} l_{0})}{I_0(p_{\beta} l_{0})} - 
F_2(R)\Big]^{-1}, \,\,\, 
D_{\beta} = -F_1 - F_2 C_{\beta} \quad,
\eeq
where
\begin{eqnarray}
\label{efy}
F_1(R)=\frac{h_1+gm_{\beta 0 }}{Kp_{\beta}I_1(p_{\beta}R)+gI_0(p_{\beta} R)}, 
\nonumber \\ \\
F_2(R)=\frac{-Kp_{\beta}K_1(p_{\beta}R)+gK_0(p_{\beta}R)}{Kp_{\beta}I_1(p_{\beta}R)+
gI_0(p_{\beta}R)} \quad. \nonumber
\end{eqnarray}

When the radially symmetric constrained magnetization is inserted into Eq.(\ref{LGW}) 
one obtains - after subtracting the bulk terms - the following expression 
for the effective interface Hamiltonian 
\beq
H[\widetilde{m}_{0}] = 2 \pi\,V(R,l_{0})
\eeq
where the function $V(R,l_{0})$ can be split into three terms
\beq
\label{V1}
V(R,l_{0}) = R\,\Sigma_{s \beta}(R) \,+\,l_{0}\,\Sigma_{\alpha \beta}(R,l_{0}) 
\,+\,R\,\omega(R,l_{0}) \quad. 
\eeq
The full expression for $V(R,l_0)$ is presented in the Appendix.

The quantity $\Sigma_{s \beta}(R)$ represents the $R$-dependent substrate-phase 
$\beta$ 
surface tension, $\Sigma_{\alpha \beta}(R,l_{0})$ is the $R$ and $l_{0}$-dependent 
interfacial stiffnes coefficient, and $\omega(R,l_{0})$ - the effective interface potential. 
The surface tension $\Sigma_{s \beta}(R)$ can be independently obtained by considering 
the disclike substrate immersed in the phase $\beta$ and the explicit expression 
for it is presented in the Appendix. 
The radially symmetric equilibrium value of $l_{0}$ will be denoted by $\bar{l}$. 
It minimizes the function $V(R,l_{0})$, i.e., 
$\frac{\partial V(R,l_{0})}{\partial l_{0}} |_{l_{0}=\bar{l}}\,=\,0$.

The next step is to find the constrained 
magnetization profile $\widetilde{m}(r,\phi,l)$ in the case when small 
deviations from the 
radially symmetric case are present, i.e.,  
$\widetilde{m}(r,\phi,l) = \widetilde{m}_{0}(r,l_{0}) + 
\delta \widetilde{m}(r,\phi,l)$ 
and $l(\phi) = l_{0} + \delta l(\phi)$. 
This can be done along the lines described in Ref.\cite{Jin}. Within 
this approach 
$\delta \widetilde{m}(r,\phi,l)=\frac{\partial 
\widetilde{m}_{0}(R,l_{0})}{\partial l_{0}} \delta l(\phi)
(1+O(\delta l^2))$. 
 We are interested in the effective interface 
Hamiltonian evaluated relative to the radially symmetric equilibrium 
configuration in which the interface has position $r=\bar{l}$.  
Up to terms quadratic in $d \delta l/ d \phi$ it takes the form 
\beq
\label{HFl8}
H_{Fl}\big[\delta l \big] \,= \,
\int d\phi \,\Big[\frac{1}{2}\frac{\Sigma_{\alpha\beta}(R,l)}
{l}\Big(\frac{d\delta l}{d\phi}\Big)^2 + V(R,l) - V(R,\bar{l}) \Big] \quad, 
\eeq
while the interfacial stiffness coefficient is identified as 
\beq
\Sigma_{\al\be}(R,l)=Kl\int dr\frac{1}{r}\Big(\frac{\partial 
\widetilde{m}_{0}(r,l)}{\partial l}\Big)^2.
\eeq

To proceed analytically we limit our considerations to the case of 
large substrate's radius $p_{\gamma}R\gg 1$, and to such choices of 
the thermodynamic 
state that $p_{\gamma} (l-R) \gg 1$ and $(l-R)/R \ll 1$. 
Then the dominant contributions to the above functions take the form 
\begin{eqnarray}
\label{ssig1}
\Sigma_{\alpha \beta}(R,l) = \sigma \left[1+O\Big(\frac{1}
{p_\alpha l}\Big) + O\Big(\frac{1}
{p_\beta l}\Big)\right] + a \tau\Big[1+O\Big(\frac{1}
{p_\beta R}\Big)\Big] \sqrt{\frac{R}{l}} \, e^{-p_\beta (l-R)} + 
\nonumber \\
+\left[ s_{20}\Big[1+O\Big(\frac{1}{p_\beta R}\Big)\Big] 
\frac{l^2}{R^2} -  2\,c\,\Big[1+O\Big(\frac{1}{p_\beta R}\Big)\Big]\,
\frac{l}{R}(l-R)   \right.  \nonumber \\ + \left.
2\,b\,\tau^2\,\frac{R}{l}\Big(1+O\frac{1}{p_\beta R}\Big) 
\right] e^{-2p_\beta(l-R)}\, +  O\left( (l-R)e^{-3p_\beta(l-R)}\right) \quad,
\end{eqnarray}

\begin{eqnarray}
\label{omeg}
l\,\Sigma_{\al \be}(R,l) + R\,\omega (R,l) = \sigma l\Big[1+O\Big(\frac{1}
{p_\beta l }\Big)\Big] + a\tau\Big[1+O\Big(\frac{1}{p_{\beta}R}\Big)\Big]
\sqrt{l R}e^{-p_{\beta}(l-R)} + \nonumber \\
(b\tau^2 + \frac{cl}{R})\Big[1+O\Big(\frac{1}{p_{\beta}R}\Big)\Big]e^{-2p_{\beta}(l-R)} 
+ O\Big[\tau\sqrt{\frac{l}{R}}\Big(1+O(\frac{1}{p_{\beta}R})\Big) 
e^{-3p_{\beta}(l-R)}\Big] \quad.
\end{eqnarray}
The coefficients $\sigma$, $a$, $b$ and $c$ are expressed in terms of the 
L-G-W Hamiltonian parametres: 
\bdm
\sigma = \frac{K}{2}(m_{\alpha 0}^2p_{\alpha}+m_{\beta 0}^2p_{\beta}),\quad
a=2Kp_{\be}m_{\be 0 }, \quad
b=Kp_{\be}, \quad
\edm \bdm 
c=Kp_{\be}\mathcal{G}m_{\be 0}^2, \quad
s_{20}=\frac{1}{2}Kp_\beta (\mathcal{G}^2+6\mathcal{G}-1) m_{\beta 0 }^2 \quad ,
\edm
where $\mathcal{G}=-\frac{Kp_{\be}+g}{Kp_{\be}-g}$, and the parameter 
$\tau = \frac{h_1+gm_{\be 0}}{Kp_{\be}-g}$ 
is interpreted as the distance from the planar substrate's wetting 
temperature ($\tau\sim \frac{T-T_W}{T_W})$. The positive values $\tau>0$ correspond to 
the wet state of the reference planar substrate. The expressions (\ref{ssig1}) and (\ref{omeg})  
reduce to those known for the planar case (see Ref.\cite{Jin}) upon neglecting 
terms proportional to powers of $(l-R)/R$ as compared to one. 

As already remarked we are interested in system's states characterized 
by large values of $R$, and $T$ close $T_{W}$. In the following we shall apply mean field approximation within which the equilibrium state minimizes $V(R,l_0)$. Fluctuation effects missed in this kind of description are known to influence quantitatively the system's critical properties. For this reason our results will not give the correct indices in the planar limit \cite{Abraham}. They will also differ from those obtained in \cite{Gil}, where the disc wetting phenomena were discussed within transfer matrix approach but no curvature modifications of the stiffness coefficient and the potential $V$ were considered. We believe that the model discussed here correctly describes the curvature effects on the quantitative level.  

Consequently, the equilibrium  
thickness of the adsorbed layer $\lambda=\bar{l}-R$, fulfils the following equation 
\begin{eqnarray}
\label{rownanie}
\frac{\sigma}{R}+\frac{a\tau}{2R}\Big(1+\frac{\lambda}{R}\Big)^{-1/2}e^{-p_{\beta}\lambda}-
p_{\beta}a\tau\Big(1+\frac{\lambda}{R}\Big)^{1/2}e^{-p_{\beta}\lambda}+ \\ \nonumber 
\Big[\frac{c}{R}-
2p_{\beta}\Big(b\tau^2+c+c\frac{\lambda}{R}\Big)\Big]e^{-2p_{\beta}\lambda}=0 \quad.
\end{eqnarray}
Its solution $\lambda$ can be determined approximately by considering the relevant case 
$\lambda\ll R$. The lowest order approximation $\lambda_{0}$ is obtained by neglecting 
in Eq.(\ref{rownanie}) all terms proportional to $\frac{\lambda}{R} $ :
\beq
\frac{\sigma}{R}-p_{\beta}a\tau e^{-p_{\beta}\lambda_0} - 
2p_{\beta}(b\tau^2+c)e^{-2p_{\beta}\lambda_0}=0  
\eeq
which leads to 
\beq
\label{l0}
p_{\be}\lambda_0= -\log\frac{-a\tau+\sqrt{a^2\tau^2+\frac{8\sigma(b\tau^2+c)}
{p_{\be}R}}}{4(b\tau^2+c)} \quad.
\eeq 
For $R\to\infty$ the above expression reduces to the well known solution for the planar case: 
$p_{\be}\lambda_0(R = \infty)= -\log\frac{-a\tau}{2(b\tau^2+c)}$. The approximation 
leading to Eq.(\ref{l0}) 
neglects all the curvature corrections to $V(R,l)$ as compared to the planar case. 
To explore 
the curvature-induced effects we solve Eq.(\ref{rownanie}) replacing  
the expressions $\frac{\lambda}{R}$ 
with $\frac{\lambda_0}{R}$. In this way the following 
equation for $\lambda_1$ is obtained 
\begin{eqnarray}
\label{rownanie1}
\frac{\sigma}{R}+\frac{a\tau}{2R}(1+\frac{\lambda_0}{R})^{-1/2}e^{-p_{\beta}\lambda_1}-
p_{\beta}a\tau\Big(1+\frac{\lambda_0}{R}\Big)^{1/2}e^{-p_{\beta}\lambda_1} + 
\\ \nonumber 
\Big[\frac{c}{R}-2p_{\beta}\Big(b\tau^2+c+c\frac{\lambda_0}{R}\Big)\Big]
e^{-2p_{\beta}\lambda_1}=0 
\end{eqnarray}
with solution 
\beq
\label{l1}
p_{\beta}\lambda_1=-\log\frac{-a\tau(1+\frac{\lambda_0}{R})^{\frac{1}{2}} + 
\sqrt{a^2\tau^2(1+\frac{\lambda_0}{R})+
\frac{8\sigma}{p_{\beta}R}(b\tau^2+c+c\frac{\lambda_0}{R})}}{4(b\tau^2 + 
c[1+\frac{\lambda_0}{R}])} \quad. 
\eeq
By iterating this procedure one defines a sequence of approximations to 
$\lambda$  denoted $\lambda_n$, $n=0,1,2, \dots$. Our analysis of 
the correlation function will be limited to $n=1$. Solving numerically 
Eq.(\ref{rownanie}) and comparing the numerical result with the approximate 
solution $\lambda_1$ shows that the difference 
$|(\lambda_1 - \lambda)/\lambda|$ does not exceed 
$10^{-6}$ (see \cite{Num}). 

It follows from Eq.(\ref{l1}) that the critical wetting state at which $\lambda$ diverges 
cannot be reached 
for finite $R$ values. This confirms the well known fact of non-existence of wetting 
transitions for surfaces with positive curvature. 

\section{Correlation function}
\renewcommand{\theequation}{3.\arabic{equation}} 
\setcounter{equation}{0}
\vspace*{0.5cm}

The interfacial correlation function is defined as 
\beq
\label{funkor}
<\delta l(\phi_1 )\,\delta l(\phi_2 )> \,= \, 
\frac{\int\mathcal{D}\delta l \,\delta l(\phi_1)\,\delta l(\phi_2)\, 
e^{-\beta H_{Fl}[\delta l]}}{\int \mathcal{D}\delta l \, 
e^{-\beta H_{Fl}[\delta l]}} \quad.
\eeq
For small gradients of the fluctuating 
interface and for small variations of its position relative to its equilibrium 
 value  
$\delta l(\phi) = l(\phi) - \bar{l}$ one may approximate the Hamiltonian, 
Eq.(\ref{HFl8}) by the expression bilinear in $\delta l$
\beq
\label{HFl2}
H_{Fl}\big[\delta l \big] = 
\frac{1}{2}\int d\phi\Big[\frac{\Sigma_{\al\be}(R,\bar{l})}{\bar{l}}
\Big(\frac{d\delta l}{d\phi}\Big)^2 + 
V^{''}(R,\bar{l}) (\delta l)^2\Big],
\eeq
where $V''(R,\bar{l}) = \left.\frac{\partial^2V}{\partial l^2}\right|_{l=\bar{l}}$.

After applying the Fourier expansion to $\delta l(\phi)$ 
\beq
\delta l(\phi)=\sum_{n=0}^{\infty}[A_n\cos n\phi +B_n\sin n\phi],
\eeq
Eq.(\ref{HFl2}) can be rewritten as 
\beq
\label{HFl3}
H_{Fl}[\delta l]=\frac{1}{4}\sum_{n=1}^{\infty}\Big(A_n^2+B_n^2\Big)
\Big(n^2\frac{\Sigma_{\al\be}(R,\bar{l})}{\bar{l}}+
V^{''}(R,\bar{l})\Big) + \frac{1}{2}A_0^2V^{''}(R,\bar{l}) \quad.
\eeq
Inserting Eq.(\ref{HFl3}) into Eq.(\ref{funkor}),  and integrating over 
$A_{0}, \{A_{n},B_{n}\}$ leads to the 
following expression for correlation function
\beq
\label{fkorelacji}
<\delta l(\phi_1)\delta l(\phi_2)>\,=\,{\cal{A}}(R,\bar{l})\,\frac{\cosh\left[\frac{1}
{\xi_{\phi}}(1-|\Delta\phi|/\pi)\right]}{\sinh(\frac{1}{\xi_{\phi}})},
\eeq
where $\Delta\phi=\phi_1-\phi_2$, $\xi_{\phi}=\pi^{-1}
\sqrt{\frac{\Sigma_{\al\be}(R,\bar{l})}
{V^{''}(R,\bar{l})\bar{l}}} $ is the correlation angle, and 
${\cal{A}}=\frac{1}{\beta V^{''}(R,\bar{l})}\frac{1}{\xi_{\phi}}$ - the correlation 
function's amplitude. The correlation function depends on the absolute value 
of the relative angle $\Delta\phi$ and is minimal at $\Delta\phi = \pi$. On the other hand, 
for the case $\Delta\phi = 0$ one obtains 
$W^2=<\delta l(\phi_1)\delta l(\phi_1)>={\cal{A}}\,\coth(1/\xi_{\phi})$. 
In the planar 
limit $R \rightarrow \infty$ the quantity $\xi_{\phi}$ corresponds to the 
parallel correlation length $\xi_{||}$ divided by $R$. \\

For further investigation of Eq.(\ref{fkorelacji}) one needs to specify 
$\bar{l}$ and 
$V^{''}(R,\bar{l})$. In what follows we shall use 
the results of Section 2 by putting $\bar{l}=R+\lambda_1$. In this way one  
obtains 
\beq
\label{omegabis}
V^{''}(\bar{l},R)=2\pi\Big[p_\beta ^2 a\tau\Big(R+\frac{\lambda_1}{2}\Big)
e^{-p_{\be}\lambda_1}+4p_{\be}^2\big[ (b\tau^2+c)R+c\lambda_1\big]e^{-2p_{\be}\lambda_1}\Big], 
\eeq
where the higher order terms have been neglected. The quantity $\Sigma_{\al\be}(R,\bar{l})$ 
is evaluated by substituting $\lambda_1$ into Eq.(\ref{ssig1}).

\section{Scaling regimes}
\renewcommand{\theequation}{4.\arabic{equation}} 
\setcounter{equation}{0}
\vspace*{0.5cm}

In this section we discuss the scaling behaviour of the correlation angle 
$\xi_{\phi}$ and the amplitude $\mathcal{A}$ as 
functions of the variables $\tau$ and $R$ in the regime where $|\tau|\ll 1$ and 
$p_{\gamma}R \gg 1$. 
We introduce the dimensionless scaling variables $t$ and $r$ 
\beq
\label{Variables}
t=\frac{a\,\tau}{4\,c},       \quad    r=\frac{a^2\,\tau^2\,p_{\be}R}{8\,\sigma\,c} 
\eeq
such that  Eq.(\ref{l0}) takes the following form 
\beq
\label{lb}
p_{\be}\lambda_0 = - \log\Big[-t+\sqrt{t^2+\frac{t^2}{r}}\Big].
\eeq
Note that the variable $r$ depending both on the  magnitude of 
$\tau $ and $1 / p_{\beta}R$  may cover arbitrary region of the 
positive semiaxis, whilst $p_\gamma R\gg 1$ implies $r/t^2\gg 1$. \\

The leading asymptotic behaviour of the correlation angle $\xi_{\phi}$ 
and the amplitude 
${\cal{A}}$ is determined by $\lambda_0$. In this case one has to 
distinguish four regimes 
in the section of the $R - \tau$ plane such that $p_{\beta} R \gg 1$ and $|\tau| \ll 1$ 
depending whether  $T > T_{W}$, $T < T_{W}$, $r \gg 1$, and $r \ll 1$. 
However, introducing $\lambda_1$ instead of $\lambda_0$, as we do, 
modifies the correlation angle $\xi_{\phi}$ and the amplitude $\mathcal{A}$ on the level of 
corrections to the leading asymptotic 
behaviour. It turns out that from this point of view eight subsets of the 
asymptotic 
regime $p_{\beta} R \gg 1$ and  $|t| \ll 1$ must be distinguished. They are 
listed below 
together with the corresponding expressions for $\lambda_1$,  $\xi_{\phi}$, 
and ${\cal{A}}$: 

\begin{enumerate}

\item $T < T_{W}$,\,$r\gg 1$,\,$|rt^2\log(-2t)|\ll 1$   \\
\beq
\left\{ \begin{array}{ll}
p_{\beta}\lambda_1\,=\, -\log(-2t)(1\,+\,\frac{1}{4r\log(-2t)}\,+\,H.O.T.) \\

\xi_{\phi}\,=\, -\frac{1}{2\sqrt{2}}\pi^{-3/2}\sqrt{\frac{2c}{\sigma}}
\frac{t}{r}\Big(1-\frac{3}{8}\frac{1}{r}\,+\,H.O.T.\Big) \\

\beta \sigma p_{\beta} {\cal{A}}= \frac{-\sqrt{2}}{4}\sqrt{\pi \frac{\sigma}{2c}}
\frac{1}{t}\Big(1-\frac{3}{8}\frac{1}{r}\,+\,H.O.T.\Big).

\end{array} \right.
\eeq

\item $T < T_{W}$, $r\gg 1$, $|rt^2\log(-2t)|\gg 1$   

\beq
\left\{ \begin{array}{ll}
p_{\beta}\lambda_1\,=\, -\log(-2t)(1\,+\,\frac{1}{4r\log(-2t)}\,+\,H.O.T.) \\

\xi_{\phi}= -\frac{1}{2\sqrt{2}}\pi^{-3/2}\sqrt{\frac{2c}{\sigma}}
\frac{t}{r}\Big(1+kt^2\log(-2t)\,+\,H.O.T.\Big) \\

\beta \sigma p_{\beta}{\cal{A}}= \frac{-\sqrt{2}}{4}\sqrt{\pi \frac{\sigma}{2c}}\frac{1}{t}
\Big(1+kt^2\log(-2t)\,+\,H.O.T.\Big),

\end{array} \right.
\eeq
where $k=-4\frac{c}{\sigma}$.

\item $T<T_W$, $r\ll 1$, $r^{3/2}\gg |t^2\log\frac{-t}{\sqrt{r}}|$

\beq
\left\{ \begin{array}{ll}
p_{\beta}\lambda_1 = -\log\frac{-t}{\sqrt{r}}(1\,+
\,\sqrt{r}\frac{1}{\log(-t/\sqrt{r})} \,+\,H.O.T.)  \\

\xi_{\phi}\,=\, -\frac{1}{2}\pi^{-3/2}\sqrt{\frac{2c}{\sigma}}
\frac{t}{\sqrt{r}}\Big(1-\frac{1}{2}\sqrt{r}\,+\,H.O.T.\Big)  \\

\beta \sigma p_{\beta}{\cal{A}}= -\frac{1}{2}\sqrt{\pi\frac{\sigma}{2c}}
\frac{\sqrt{r}}{t}\Big(1-\frac{1}{2}\sqrt{r}\,+\,H.O.T.\Big).

\end{array} \right.
\eeq

\item $T<T_W$, $r\ll 1$, $r^{3/2} \ll |t^2\log\frac{-t}{\sqrt{r}}|$

\beq
\left\{ \begin{array}{ll}
p_{\beta}\lambda_1 \,=\,-\log\frac{-t}{\sqrt{r}}(1+\frac{c}{\sigma}\frac{t^2}{r}\,+\,H.O.T.)\\
\xi_{\phi}\,=\, -\frac{1}{2}\pi^{-3/2}\sqrt{\frac{2c}{\sigma}}\frac{t}{\sqrt{r}}
\Big(1+k\frac{t^2}{r}\log\frac{-t}{\sqrt{r}}\,+\,H.O.T.\Big)  \\
\beta \sigma p_{\beta}\mathcal{A} \,=\, -\frac{1}{2}\sqrt{\pi\frac{\sigma}{2c}}\frac{\sqrt{r}}{t}
\Big(1+k\frac{t^2}{r}\log\frac{-t}{\sqrt{r}}\,+\,H.O.T.\Big).
\end{array} \right.
\eeq
\item $T>T_W$, $r\gg 1$,  $|t^2\log\frac{t}{2r}|\ll 1$
\beq
\left\{ \begin{array}{ll}
p_{\beta}\lambda_1\,=\, -\log\frac{t}{2r}(1-\frac{1}{2r\log (t/2r)}+H.O.T.)  \\

\xi_{\phi}=\frac{1}{\sqrt{2}}\pi^{-3/2}\sqrt{\frac{2c}{\sigma}}
\frac{t}{\sqrt{r}}\Big(1+\frac{3c}{\sigma}\frac{t^2}{r}\log\frac{t}{2r}+H.O.T.\Big)   \\

\beta \sigma p_{\beta}{\cal{A}}=\frac{\sqrt{2}}{2}\sqrt{\pi \frac{\sigma}{2c}}
\frac{\sqrt{r}}{t}\Big(1-\frac{2c}{\sigma}\frac{t^2}{r}\log\frac{t}{2r}+H.O.T.\Big).
\end{array} \right.
\eeq
\item $T>T_W$, $r \gg 1$, $|t^2\log\frac{t}{2r}|\gg 1$
\beq
\left\{ \begin{array}{ll}
p_{\beta}\lambda_1= -\log\frac{t}{2r}(1-2\frac{c}{\sigma}\frac{t^2}{r}+H.O.T.)  \\

\xi_{\phi}= \frac{1}{\sqrt{2}}\pi^{-3/2}\sqrt{\frac{2c}{\sigma}}
\frac{t}{\sqrt{r}}\Big(1+\frac{5}{2}\frac{c}{\sigma}\frac{t^2}{r}\log\frac{t}{2r}+H.O.T.\Big) \\

\beta \sigma p_{\beta}{\cal{A}} = \frac{\sqrt{2}}{2}\sqrt{\pi \frac{\sigma}{2\,c}}
\frac{\sqrt{r}}{t}\Big(1+\frac{c}{2\sigma}\frac{t^2}{r}\log\frac{t}{2r}+H.O.T.\Big).
\end{array} \right.
\eeq
\item $T>T_W$, $r\ll 1$, $r^{3/2}\gg |t^2\log\frac{t}{\sqrt{r}}|$
\beq
\left\{ \begin{array}{ll}
p_{\beta}\lambda_1= -\log\frac{t}{\sqrt{r}}(1-\frac{\sqrt{r}}{\log(t/\sqrt{r})}+H.O.T.)  \\

\xi_{\phi}=\frac{1}{2}\pi^{-3/2}\sqrt{\frac{2c}{\sigma}}
\frac{t}{\sqrt{r}}\Big(1+\frac{1}{2}\sqrt{r}+H.O.T.\Big)   \\

\beta \sigma p_{\beta}{\cal{A}}=\frac{1}{2}\sqrt{\pi \frac{\sigma}{2c}}
\frac{\sqrt{r}}{t}\Big(1+\frac{1}{2}\sqrt{r}+H.O.T.\Big).
\end{array} \right.
\eeq
\item $T>T_W$, $r\ll 1$, $r^{3/2}\ll |t^2\log\frac{t}{\sqrt{r}}|$
\beq
\left\{ \begin{array}{ll}
p_{\beta}\lambda_1= -\log\frac{t}{\sqrt{r}}(1+\frac{c}{\sigma}\frac{t^2}{r}+H.O.T.)  \\

\xi_{\phi}=\frac{1}{2}\pi^{-3/2}\sqrt{\frac{2c}{\sigma}}
\frac{t}{\sqrt{r}}\Big(1+k\frac{t^2}{r}\log\frac{t}{\sqrt{r}}+H.O.T.\Big)    \\

\beta \sigma p_{\beta}{\cal{A}}=\frac{1}{2}\sqrt{\pi \frac{\sigma}{2c}}
\frac{\sqrt{r}}{t}\Big(1+k\frac{t^2}{r}\log\frac{t}{\sqrt{r}}+H.O.T.\Big).
\end{array} \right.
\eeq
\end{enumerate}

In all the specified regimes of parameters $R$ and $\tau$, the quantity $\lambda_1$ 
diverges logarithmically either as function of $\tau$ or as function of $R$. 
In the above regimes 1 and 2 the leading term is only temperature dependent, i.e., 
$p_{\beta}\,\lambda_1 \sim - \ln|\tau|$ whilst for the cases 3, 4, 7 and 8 it depends 
exclusively on the radius, i.e.,  $p_{\beta}\,\lambda_1 \sim \ln(p_{\beta}R)$. 
The leading contribution to the correlation angle is proportional to 
$\frac{1}{R\tau}$ for regimes 1 and 2, which for the corresponding correlation length 
$\xi_{||}=R\xi_\phi$ gives $\xi_{||}\sim\frac{1}{|\tau|}$. On the other hand, 
for cases 3-8 we obtain $\xi_\phi\sim R^{-1/2}$, which means that the dominant 
term is temperature-independent and converges to $0$ in the limit $R\to\infty$. 
In these cases one has $\xi_{||}\sim R^{1/2}$. 
For the amplitude ${\cal{A}}$ we obtain ${\cal{A}}\sim\tau^{-1}$ in cases 1, 2, and 
${\cal{A}}\sim R^{1/2}$ 
in regimes 3-8. Note that in regimes 3-8 the interfacial width $W$ becomes proportional 
to $ R^{1/4}$. 
We also observe that the dependence of the leading term to the derived expressions for 
$\xi_\phi$  and ${\cal{A}}$ on 
the free $\alpha-\beta$ interfacial stiffness coefficient $\sigma$ is not the same 
for all the above specified cases. In particular, at fixed $R$ and $\tau$ 
in regions 1 and 2 the quantity $\xi_\phi$ is proportional to $\sqrt\sigma$ 
while in regimes 3-8 the leading contribution to $\xi_\phi$ does not depend on 
$\sigma$. 
On the other hand, as far as the dominant contribution is concerned,  
in regions 1 and 2 the amplitude ${\cal{A}} \sim 1/\sqrt\sigma$ while in all the other cases
 -  ${\cal{A}} \sim 1/\sigma$.

Among the derived corrections one encounters terms which are nonanalytic 
in the zero curvature limit when the temperature approaches the wetting temperature 
of the reference planar system. (For discussion of non-analytic contributions 
to other thermodynamic quantities in wetting systems, see eg. 
Refs.\cite{Holyst}, \cite{Evans}).  These terms contain a logarithm of $R$ or $\tau$ 
multiplied by a power of $1/R$ or $\tau$, which is characteristic for the considered 
potential $V(R,l)$, appropriate for short-range interactions.  In our 
treatment of the correlation function the presence of the non-analytic terms is 
a consequence of the structure of the potential $V(R,l)$ and the stiffness 
coefficient $\Sigma_{\al\be}(R,l)$ derived in Section 2. The form of corrections to the 
leading asymptotic behavior of $\xi_{\phi}$ and $\mathcal{A}$ follows from the 
curvature-dependent potential $V(R,l)$ in regimes 1, 3, 5, 6, and 7. In regimes 2, 4, 
and 8 their form follows exclusively from the necessity of incorporating the 
position dependence of the stiffness coefficient $\Sigma_{\al\be} (R,l)$ into the analysis. 
To consider still the next to leading order corrections $\xi_{\phi}$ and ${\cal{A}}$ one 
should use - within the present approach - a better approximation to 
$\lambda$ (say, $\lambda_2$ instead of $\lambda_1$). However, additional restrictions must 
then be imposed on the parametres $\tau$ and $R$ to determine the character of 
the next correction; accounting for it causes necessity to 
increase the number of regimes of parametres $\tau$ and $R$ that must be distinguished. 

\section{Summary}
\renewcommand{\theequation}{5.\arabic{equation}} 
\setcounter{equation}{0}
\vspace*{0.5cm}

Starting from the Landau-Ginzburg-Wilson theory we derived within the mean field 
approximation 
the effective Hamiltonian for the fluctuations of an interface separating two 
coexisting phases,  
one of them forming a layer adsorbed on a disclike  substrate. The Hamiltonian is 
parametrized by the temperature $T$ and the disc radius $R$, i.e., both the coefficient 
of the interfacial stiffness  $\Sigma_{\alpha\beta}$ and the interface potential $V$ 
appearing in the effective Hamiltonian depend 
on these two parameters, and on the interface position $l$.   
We applied the Gaussian approximation and found 
the expression for the height-height correlation function, in particular the correlation angle 
$\xi_{\phi}$ and the amplitude ${\cal{A}}$. 
Considering the case of large disc radii and temperatures close to the wetting 
temperature $T_W$ of a reference planar system and keeping track of the position 
dependence of the stiffness coefficient $\Sigma_{\alpha\beta}$ and the 
curvature corrections to the effective interaction potential $V$ we identified the scaling 
regimes in the space of parametres $R$ and $T-T_W$ in which the dominant contributions 
to  $\xi_{\phi}$ and ${\cal{A}}$ were determined. In addition we have derived the expression for the 
$R$-dependent surface tension coefficient $\Sigma_{s\beta}$ characterizing the 
substrate-phase $\beta$ interface. This surface tension coefficient can be represented as 
the power series in $1/R$, and the corresponding Tolman's length \cite{Tolman},\cite{Henderson} turns out to be 
proportional  to $(T-T_{W})^2$. 

In the limit of large disc radii our results for the equilibrium adsorbate layer thickness
$\lambda$ and the correlation length $\xi_{||} \sim R \xi_{\phi}$ differ from 
those obtained 
by Gil and Mikheev (G-M) in Ref.\cite{Gil} via the transfer matrix approach. 
These authors' analysis referred to a temperature independent steplike potential $V_S(\lambda)/R=0$ for $\lambda >0$, $V_S(\lambda)/R=\infty$ for $\lambda\leq 0$ which captures the relevant features of complete wetting phenomena in the planar limit. For that case they predict  $\bar{l}\sim R^{1/3}$ and $\xi_{||}\sim R^{2/3}$.  The discrepancy between our and G-M result is due to fluctuation effects missed in our description. For the same reason we fail to predict the correct values of the critical indices in the planar 2D limit \cite{Abraham}. Performing a transfer matrix calculation taking into account the effect of curvature on the structure of the effective Hamiltonian is a demanding task. We believe that the MFT approach applied here generally accounts for the observable phenomena, though obviously the system's behaviour is further modified when fluctuations are incorporated into the description. 
It is worth noting that the   
phenomenological treatment based on balancing the curvature-induced Laplace pressure 
and the disjoining pressure applied by G-M to the case of long range 
intractions yields in the case of short range forces the results consistent with ours. 

From this perspective it would be very interesting to analyze the interfacial correlation 
function in 
system with short range forces by properly taking into account all the relevant fluctuations. 
It has been shown  \cite{Jin,Jin2} that such planar systems may undergo either first or second 
order wetting depending on the system parameters. However, this alternative applies to systems 
with dimension larger than $d_c=2.41$, \cite{Boulter1}, which is not the case considered in this paper. 
On the other hand, whether the interfacial correlations in the case of 
cylinder- or spherelike substrates contain some indications of the possible  
change of the order of the wetting transion in the corresponding planar case remains an open question.\\ 

\noindent {\bf {Acknowledgment}}  \\ 
\noindent This work is supported by the Committee for Scientific Research via 
grant 2PO3B 008 23.

\section{Appendix}
\renewcommand{\theequation}{A.\arabic{equation}} 
\setcounter{equation}{0}
\vspace*{0.5cm}
In this Appendix we quote the complete expressions for the effective interface 
Hamiltonian $V(R,l)$ in 
the radially symmetric case and for the substrate-phase $\beta$ surface tension 
$\Sigma_{s \beta}(R)$ 
as derived from the L-G-W theory. These expressions have the following form:
\begin{displaymath}
V(R,l)\, =\, \frac{K}{2} \Bigg[\frac{1}{2}C_{\beta}^2p_\beta^2\Big[l^2K_0(p_\beta l)
\Big(K_0(p_\beta l)-K_2(p_\beta l)\Big)-
R^2K_0(p_\beta R)\times
\end{displaymath}
\begin{displaymath}
\times\Big(K_0(p_\beta R)-K_2(p_\beta R)\Big)\Big]+\frac{1}{2}D_\beta^2p_\beta^2
\Big[l^2I_0(p_\beta l)\Big(I_0(p_\beta l)-
I_2(p_\beta l)\Big)-R^2I_0(p_\beta R)\times
\end {displaymath}
\begin{displaymath}
\times\Big(I_0(p_\beta R)-I_2(p_\beta R)\Big)\Big]+
2C_\beta D_\beta\int_{p_\beta R}^{p_\beta l}[K_0(u)I_0(u)-
K_1(u)I_1(u)]udu\Bigg]+
\end{displaymath}
\begin{displaymath}
+2\pi R\Bigg[-h_1m_{\beta 0h}-\frac{1}{2}gm_{\beta 0h}^2-
\Big(h_1+gm_{\beta 0h}\Big)C_\beta K_0(p_\beta R)-\Big(h_1+gm_{\beta 
0h}\Big)D_\beta I_0(p_\beta R)+
\end{displaymath}
\begin{displaymath}
-\frac{1}{2}gC_\beta^2K_0^2(p_\beta R)-\frac{1}{2}gD_\beta^2I_0^2(p_\beta R)-
gC_\beta D_\beta K_0(p_\beta R)I_0(p_\beta 
R)\Bigg]+
\end{displaymath}
\beq
+\frac{1}{2}K\pi m_{\alpha 0h}^2p_\alpha^2l^2\Big[\frac{K_2(p_\alpha l)}{K_0(p_\alpha l)}-1\Big] 
\nonumber \quad,
\eeq
and 
\begin{eqnarray}
\Sigma_{s\beta}(R) = - h_1\,m_{\beta 0h} - \frac{1}{2}g\,m_{\beta 0h}^2 + \nonumber \\
\frac{K}{4}\,\Big(\frac{F_1(R)}{F_2(R)}\Big)^2\,p_{\beta}^2R\,K_0(p_{\beta}R)
\Big[K_2(p_{\beta}R)-K_0(p_{\beta}R)\Big] + \nonumber \\ 
-\frac{1}{2}g\, \Big(\frac{F_1(R)}{F_2(R)}\Big)^2\,K_0^2(p_{\beta}R) + \Big(h_1+ 
gm_{\beta 0h}\Big)\frac{F_1(R)}{F_2(R)}K_0(p_\beta R)  \quad,
\end{eqnarray}
where the functions $F_{1}(R)$ and $F_{2}(R)$ are defined in the main text, 
see Eq.(\ref{efy}). 
The integral $\int_{p_\beta R}^{p_\beta l}[K_0(u)I_0(u)-K_1(u)I_1(u)]du$ in the 
above expression for $V(R,l)$ 
can be performed analytically for the case $p_\beta R\gg 1$, $p_\beta l\gg1$ by 
applying the 
asymptotic expansion of the Bessel functions. The asymptotic expansion of this 
formula is quoted 
in the main text, see Eqs(\ref{ssig1}), (\ref{omeg}). \\ 

The expression for the surface tension 
$\Sigma_{s\beta}(R)$ can be derived by analyzing the the L-G-W 
theory for a  disc immersed in phase $\beta$. One notes that the function 
$\Sigma_{s\beta}(R)$ can 
be expanded in powers of $1/p_{\beta}R$ and the first two terms of this 
expansion have the following 
form:
\begin{eqnarray}
\Sigma_{s\beta}(R) = \frac{K\,p_{\beta}}{4}\,\tau^2\,-\,h_{1}\,(m_{\beta0}+\tau)-
\frac{g}{2}(m_{\beta0}+\tau)^2 \,+ \nonumber \\ 
\frac{K\,p_{\beta}\,\tau^2}{4}\,\,\frac{1}{p_{\beta}R} + H.O.T. \quad. 
\end{eqnarray}
Thus we observe that the quantity identified as the Tolman's length  
depends on the temperature and 
vanishes upon approching  the planar substrate wetting temperature $T_{W}$ 
proportionally to $\tau^2$. 
This behavior of the Tolman's length reflects the general property of 
$\Sigma_{s\beta}(R)$ which can be 
read off from the above formula, namely that at $T=T_{W}$ the surface tension 
ceases depending on $R$. \\

Finally, within the L-G-W theory the expression $\Sigma_{\alpha\beta}(R,l)$ 
for the $R$ and $l$ dependent 
interfacial tension is identified as 
$\Sigma_{\alpha\beta}(R,l) = K\,l\,\int_{R}^{\infty} dr \frac{1}{r} 
(\frac{\partial \widetilde{m}(r,l)}{\partial r})^2$. The integration can 
be performed in the asymptotic regime 
$p_{\gamma}R \gg 1$ and the result is quotes in the main text, see Eq.(\ref{ssig1}). \\


\begin{thebibliography}{99}
\bibitem{Num} The numerical check of the proposed approximation's accuracy was carried 
out for $\frac{a}{b}=\frac{a}{c}=\frac{a}{\sigma}=1$, $p_{\beta}R\in [10^2, 10^4]$ 
and $\tau\in [-0.5, 0.5]$.
\end{thebibliography}
\end{document}